\documentclass[aps,twocolumn]{revtex4}

\usepackage{graphicx}
\usepackage{comment}
\usepackage{amssymb}
\usepackage{subfigure}
\usepackage{stackrel}
\usepackage{enumitem}
\usepackage{amsmath}
\usepackage{amsfonts}
\usepackage{bm}
\usepackage{verbatim} 
\usepackage[normalem]{ulem}
\usepackage{soul}
\usepackage{mathtools}
\usepackage{xcolor}
\newcommand{\beq}{\begin{equation}}
\newcommand{\eeq}{\end{equation}}

\newcommand{\ket}[1]{\ensuremath{\left|{#1}\right\rangle}}

\definecolor{rosita}{rgb}{0.97, 0.56, 0.65}
\definecolor{naranja}{rgb}{1, 0.55, 0}
\definecolor{bleudefrance}{rgb}{0.19, 0.55, 0.91}

\begin{document}

\title{Optimal entanglement generation in GHZ-type states}

\author{N. Giovenale$^1$, L. Hernandez-Martinez$^2$, A. P. Majtey$^1$, and A. Valdés-Hernández$^2$}

\address{$^1$Instituto de F\'{i}sica Enrique Gaviola, CONICET and Universidad 
Nacional de C\'{o}rdoba,
Ciudad Universitaria, X5016LAE, C\'{o}rdoba, Argentina.\\$^2$ Instituto de F\'{\i}sica, Universidad Nacional Aut\'{o}noma de M\'{e}xico, Apartado Postal 20-364, Ciudad de M\'{e}xico, Mexico.}
\begin{abstract}
The entanglement production is key for many applications in the realm of quantum information, but so is the identification of processes that allow to create entanglement in a fast and sustained way. 
Most of the advances in this direction have been circumscribed to bipartite systems only, and the rate of entanglement in multipartite system has been much less explored. 
Here we contribute to the identification of processes that favor the fastest and sustained  generation of tripartite entanglement in a class of 3-qubit GHZ-type states. 
 By considering a three-party interaction Hamiltonian, we analyse the dynamics of the 3-tangle and the entanglement rate 
 to identify the optimal local operations that supplement the Hamiltonian evolution in order to speed-up the generation of three-way entanglement, and to prevent its decay below a predetermined threshold value.
The appropriate local operation that maximizes the speed at which a highly-entangled state is reached has the advantage of requiring access to only one of the qubits, yet depends on the actual state of the system. 
Other universal (state-independent) local operations are found that conform schemes to maintain a sufficiently high amount of 3-tangle. 
Our results expand our  understanding of entanglement rates to multipartite systems, and offer guidance regarding the strategies that improve the efficiency in various quantum information processing tasks.

\end{abstract}

\maketitle

\section{Introduction}


In the realm of quantum information, generating entanglement in a sustained way typically requires controlled operations that may pose significant difficulties. 
This calls for studies that can guide us in effectively harnessing interactions that facilitate the entanglement  production, and prevent such resource from being lost under certain dynamics. 
In particular, the efficient generation of \emph{multipartite} entanglement represents a relevant challenge, both theoretically and experimentally. 
It deserves attention mainly because multipartite systems can exhibit different types of entanglement that are key resources for many quantum information processing tasks. 
In 3-partite systems, specifically,
genuine tripartite entanglement generation is an active area of research, and new techniques and advancements are continuously emerging.
In recent years, efforts have been invested in exploring novel experimental platforms, refining existing techniques, and developing new theoretical frameworks to further understand and exploit the power of genuine tripartite entanglement for various applications in quantum information processing and quantum communication \cite{Mamaev2020,Neeley2010,Blasiak2019,Menke2022,Gao2005,Man2006,Hillery1999}.

Along with the goal of generating multipartite entangled states, it is desirable to generate them as fast as possible, and to prevent its eventual decay induced by the specific dynamics that govern the evolution of the system. Achieving this will impact the applicability of multipartite systems in quantum computing tasks.
The production of entanglement in the minimal possible time has been explored from various perspectives.   
One line of research that has attracted considerable attention focuses on identifying the most efficient use that can be made of a given Hamiltonian for generating entanglement, by studying the Hamiltonian entanglement capabilities and the entanglement rates. 
The entanglement rate of a non-local Hamiltonian acting on a two-qubit system was introduced in \cite{Dur2001}, where it was shown that the entangling process can be more efficient by supplementing the action of the Hamiltonian with local unitary operations or by using ancilla systems.
Entangling capabilities of unitary transformations acting on a bipartite quantum system of arbitrary dimension were studied in \cite{Zanardi2000}. 
In \cite{Lari2009}  a geometric approach to quantify the capability of creating entanglement for a general physical interaction acting on two qubits is developed. Bounds for the entanglement capabilities have been further explored in \cite{Bennett2003,Bravyi2007,Acoleyen2013,Chilsd2002,WangPRA2003,NingIJTP2016}. 
 Other related approaches attempt to answer what is the minimum time for reaching a target entangled state, leading for example to the study of the quantum braquistochrone problem \cite{Pati2012}.   
 
 Most of the aforementioned studies circumscribe to bipartite systems, hence are limited to the production of bipartite entanglement. Only in 
 \cite{Lari2009} the first steps to study  the  genuine three-qubit entanglement capability is presented. 
 In a system of three qubits, there are two inequivalent ways in which the parties can be genuinely entangled (meaning that non-vanishing correlations exists across all three bipartitions of the system) \cite{Dur2000,Amico2008,Horodecki2009,Guhne2009}. 
Two families of states are identified, according to the type of entanglement exhibited by its elements: the first family corresponds to GHZ-type states, which posses a non-zero \emph{3-tangle} \cite{Coffman2000,Mermin1990,Dur2000,Sabin2008}, indicating that a three-way entanglement is present \cite{Coffman2000}.
The second family corresponds to W-type states, characterized by having vanishing 3-tangle, so all the entanglement across a given bipartition decomposes as sums of pairwise correlations.  
The GHZ-type and the W-type states have different properties and applications in quantum information processing \cite{Gao2005,Man2006,Hillery1999}.

 In this paper we contribute to the study of fast generation of genuine entanglement in a class of 3-qubit states that pertain to the GHZ-type family. 
 By considering a three-party interaction Hamiltonian that preserves the structure of the initial state along its evolution, we analyse the dynamics of the 3-tangle and the entanglement rate to design  appropriate local operations, some of them requiring access to only one of the qubits, aimed at increasing the speed at which a highly-entangled state is reached. 
 Protocols are proposed that combine the Hamiltonian evolution with single-qubit operations to prevent the 3-tangle from decaying below a certain desired threshold value. 
 Our results offer guidance in the understanding of entanglement rates in multipartite systems, with potential impact in the implementation of more efficient quantum information processing tasks with enhanced entanglement properties \cite{Menke2022}.       

The article is structured as follows. Section \ref{dynamics} introduces the system  of interest, and provides an analysis of the Hamiltonian dynamics of the tripartite-entanglement and the entanglement rate, focusing on the extremal values of these quantities.
In Section \ref{opti} we tackle the problem of optimizing the entanglement rate by means of unitary operations that are performed on a single qubit, in combination with the Hamiltonian evolution.
Section IV is devoted to study different schemes that enable a fast generation of an entangled state while maintaining a high degree of entanglement throughout the evolution, by supplementing the Hamiltonian with local operations. Finally, in Section V, we present some final remarks.

\section{Entanglement dynamics in 3-qubit GHZ-like states} \label{dynamics}

\subsection{Hamiltonian evolution}
We explore the entanglement evolution of GHZ-like states with the following structure
\begin{equation} \label{psigral}
 \ket{\psi} =  \alpha \ket{000} + \beta  \, \ket{111}, 
\end{equation}
with $|\alpha|^2+|\beta|^2=1$, and $\{\ket{0}, \ket{1}\}$ the eigenestates of the Pauli operator $\sigma_z$ with eigenvalues $\{+,-\}$, respectively.
We will confine our attention to Hamiltonians that involve interactions among the 3 qubits \cite{Menke2022}, and preserve the structure of the state (\ref{psigral}). In particular, we focus on interaction Hamiltonians of the form 
\begin{equation}\label{H2}
H=\sum_{ijk}g_{ijk}\sigma_i\otimes\sigma_j\otimes \sigma_k,
\end{equation}
where $i,j,k\in\{x,y\}$, with $\sigma_{x,y}$ the Pauli operators. 
By an appropriate redefinition of the interaction coupling constants, and considering its action on states (\ref{psigral}),
the Hamiltonian (\ref{H2}) can be substituted by the simpler expression
\begin{equation}\label{Hamiltonian}
H=\gamma_x\,\sigma_{x} \otimes \sigma_{x} \otimes \sigma_{x} + \gamma_y\,\sigma_y \otimes \sigma_{y} \otimes \sigma_{y} ,
\end{equation}
which is the one considered in what follows.

An
initial state of the form \begin{equation} \label{eqn:1}
 \ket{\psi_0} =  \sin \phi \, \ket{000} + e^{i \varphi} \, \cos \phi \, \ket{111}, 
\end{equation}  
with $\phi\in[0,\pi/2]$ 
 and $\varphi\in[0,2\pi]$, evolves under the evolution operator $U=e^{-iH t/\hbar}$, with $H$ given by (\ref{Hamiltonian}). The state at time $t$ thus reads  
\beq\label{psit}
\ket{\psi (t)} =  e^{i \alpha_1(t)}\sqrt{p (t)} \, \ket{000} + e^{i \alpha_2(t)} \, \sqrt{1 - p (t)} \, \ket{111},
\eeq
with $0\leq p(t)\leq 1$, and $\alpha_{1,2}(t)$ real functions given by
\begin{equation} \label{eqn:6}
\tan \alpha_{1} (t)= \frac{ \tan \left( \frac{\pi}{2} \frac{t}{T} \right) \left( - r \, \cos \varphi  + \sin \varphi  \right)}{ \sqrt{r^{2}+1} \, \tan \phi  +  \tan \left( \frac{\pi}{2} \frac{t}{T} \right) \, \left( r \, \sin \varphi + \cos \varphi \right)} ,
\end{equation}
\begin{equation} \label{alpha2}
\tan \alpha_{2}(t) = \frac{ \sqrt{r^{2}+1} \, \sin \varphi - r \, \tan \left( \frac{\pi}{2} \frac{t}{T} \right) \, \tan \phi }{ \sqrt{r^{2}+1} \, \cos \varphi -  \tan \left( \frac{\pi}{2} \frac{t}{T}\right) \, \tan \phi} ,
\end{equation}
\begin{equation} \label{p(t)}
 p(t) = \sin^{2} \left( \frac{\pi}{2} \frac{t}{T} + \phi \right) - b \, \sin \left( \pi \frac{t}{T} \right) \, \sin 2 \phi ,
\end{equation}
where we defined $r=\gamma_x/\gamma_y$, and introduced a characteristic time
\beq \label{tiempoT}
T = \frac{ \pi\hbar}{2 |\gamma_y|\sqrt{r^{2}+1}} ,
\eeq
uniquely determined by the Hamiltonian parameters. The additional parameter $b$ reads 
\begin{equation} \label{eqn:10}
 b = \frac{1}{2} \, \left( 1 - \frac{r \, \sin \varphi +  \cos \varphi}{ \sqrt{r^{2}+1} } \right) ,
\end{equation}
and is determined by the Hamiltonian and the initial relative phase $\varphi$. It can be seen that, independently of $r$, this parameter takes values in the interval $[0,1]$.  

The state (\ref{psit}) is physically equivalent to $|\tilde{\psi} (t)\rangle=e^{-i\alpha_1(t)}|\psi (t)\rangle$, which has the form
\begin{eqnarray} \label{psi_equiv}
 |\tilde{\psi} (t)\rangle &=&  \sqrt{p(t)} \, \ket{000} + e^{i \varphi(t)} \, \sqrt{1-p(t)} \, \ket{111} \nonumber\\
 &=&\sin\phi(t) \, \ket{000} + e^{i \varphi(t)} \, \cos\phi(t) \, \ket{111},
\end{eqnarray}
with $\phi(t)\in[0,\pi/2]$ defined from $\sin \phi(t)\equiv\sqrt{p(t)}$,
 and $\varphi(t)=\alpha_{2}(t) - \alpha_{1}(t)$. 
 Therefore, the evolved state has always the same structure as the initial one. 

\subsection{Tripartite entanglement}\label{sec:entrel-trip}

Since we are dealing with a GHZ-type 3-qubit state, we will resort to the 3-tangle \cite{Coffman2000}
to quantify the amount of tripartite (genuine) entanglement. The 3-tangle is defined as
\begin{equation}\label{eqn:4}
    \tau=C^2_{a|bc}-C^2_{a|b}-C^2_{a|c}
\end{equation}
where $C_{i|j}$ is the concurrence between subsystems $i$ and $j$ \cite{Hill1997,Wootters1998,Rungta2001}.
For a general 3-qubit state of the form 
\begin{equation}\label{Evolvedstate}
\ket{\eta} = \sum_{n,l,m=0,1}c_{nlm}\ket{nlm},
\end{equation}
and in terms of $a_{ij}\equiv c_{0ij}$ and $b_{ij}\equiv c_{1ij}$, the 3-tangle can be directly computed from 
\begin{equation}
\tau =4\vert d_{1}-2d_{2}+4d_{3}\vert, \label{taud}
\end{equation}
where \cite{Coffman2000}
\begin{subequations}
\begin{eqnarray}
d_{1}&=& a^{2}_{00}b^{2}_{11}+a^{2}_{01}b^{2}_{10}+a^{2}_{10}b^{2}_{01}+a^{2}_{11}b^{2}_{00},\\
d_{2}&=&a_{00}a_{11}b_{00}b_{11}+a_{01}a_{10}b_{10}b_{01}+\\ &&(a_{10}b_{01}+a_{01}b_{10})(a_{00}b_{11}+a_{11}b_{00}),\nonumber\\
d_{3}&=& a_{00}a_{11}b_{10}b_{01}+a_{01}a_{10}b_{00}b_{11}\label{d3}.
\end{eqnarray}
\end{subequations}

From here and Eq. (\ref{psi_equiv}) the 3-tangle takes the simple form 
\beq\label{tau}
\tau (t) = 4 p(t) \, [1-p(t)] =\sin^2 2\phi(t). 
\eeq
The square of $\tau$ coincides, for the particular family of states considered, with the recently introduced triangle measure $\mathcal E_{\triangle}$ of 3-partite entanglement \cite{Xie2021}.
From Eq. (\ref{p(t)}) it follows that
\beq \label{tau_adim}
\tau (t) = 1-\Delta(t),
\eeq
with 
\beq
\Delta (t) = \Big[ (1-2  b) \, \sin 2 \phi\,\sin (\pi t/T)   - \cos 2 \phi\,\cos (\pi t/T) \,  \Big]^{2}.
\eeq  
%
From this last expression, it can be verified that $T$ is precisely the period of $\tau$. In what follows we will use the simplified notation $\tilde{t}$ to refer to the dimensionless time 
\begin{equation}\label{ttilde}
\tilde{t}\equiv\frac{t}{T}=\frac{2|\gamma_y|\sqrt{r^2+1}}{\pi\hbar}\,t.
\end{equation}
\paragraph{Maximal values of 3-tangle.-}

The maximum value of $\tau$ equals 1, and is reached at times $\tilde t_{\max}$ so
\beq \label{taumax}
\tau_{\max}=\tau(\tilde t_{\max})=1,
\eeq
with $\tilde t_{\max}$ determined by the positive roots of $\Delta(t)$, i.e., determined by positive solutions of the equation 
\begin{equation}
\label{cond_tmax}
(1-2 \, b) \sin 2 \phi\, \sin \pi\tilde t_{\max}\,   = \cos 2 \phi\,\cos \pi\tilde t_{\max}. 
\end{equation}

If $\cos 2\phi=0$, then $\tau(0)=1$ and the times at which this maximum value is attained again are integers multiples of the period $T$, so in this case the solutions to (\ref{cond_tmax}) are
\begin{equation}
    \tilde t_{\max}=k, \quad k=1,2,\dots.
\end{equation}
If $\sin 2\phi=0$ (meaning that $\tau(0)=0$), Eq. (\ref{cond_tmax}) reduce to $\cos \pi\tilde t_{\max}=0$, with solutions
\begin{equation}\label{solimpar}
    \tilde t_{\max}=\frac12(2k+1), \quad k=0,1,\dots.
\end{equation}
that are half integers multiples of the period $T$.

If $1-2b=0$ (with $\cos 2\phi\neq0$), the solutions to Eq. (\ref{cond_tmax}) are the same as those in (\ref{solimpar}).

For $(1-2b)\neq 0$, $\cos 2\phi\neq 0$, and $\sin 2\phi\neq 0$, we get
\begin{equation}\label{tmax}
\tilde{t}_{\max} = \frac{ 1}{\pi} \, \left\{ \arctan \left[ \frac{1}{(1-2b) \, \tan 2 \phi} \right] + k \, \pi \right\} ,
\end{equation}
with $k=0, 1,2,\dots$ suitably chosen to ensure that $\tilde{t}_{\max}>0$. 

\begin{figure}[ht]
    \centering
    \includegraphics[width=.8\linewidth]{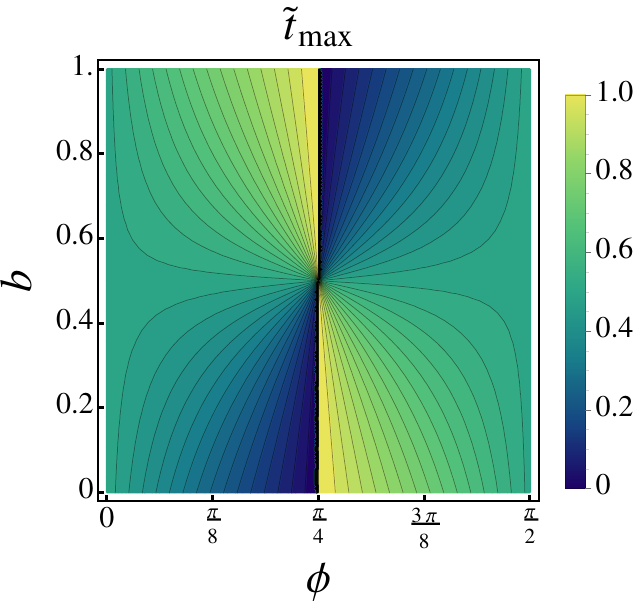}
    \caption{Time $\tilde{t}_{max}$ at which the state $\ket{\psi_0}$ reaches $\tau=1$ for the first time (lowest positive solution of Eq. (\ref{tmax})), as a function of the initial state's parameters $\phi$ and $b$.}
    \label{fig:tmax}
\end{figure}

For all initial states of the form (\ref{eqn:1}) $\tilde t_{\max}\leq T$, so \emph{all} initial states reach the maximum entanglement under the Hamiltonian evolution. 
This can be verified in Fig.~\ref{fig:tmax}, which shows the first time at which the maximal 3-tangle is attained, as a function of $b$ and $\phi$.

The particular case, which has been excluded in the above analysis, with $b=1/2$ \emph{and} $\cos 2 \phi=0$ corresponds to an initial state with
\beq \label{tauconstant}
\phi_\textrm{s}=\frac{\pi}{4}\quad \mbox{and}\quad\varphi_\textrm{s}=\arctan\Big(-\frac{1}{r}\Big)+k\pi,
\eeq
where $k=0,1\dots$, and is a maximally entangled ($\tau(t)=1$) eigenstate of the Hamiltonian (\ref{Hamiltonian}).


\paragraph{Minimal values of 3-tangle.-} The minimal value of Eq. (\ref{tau_adim}) is given by
\begin{eqnarray}\label{taumin}
\tau_{\min}&=&\tau(\tilde{t}_{\min})=4b(1-b)\sin^2\,2\phi\nonumber\\
&=&4b(1-b)\tau(0),
\end{eqnarray}
where in the last line we used (\ref{tau}) to write $\tau(0)=\sin^2\,2\phi$.
Equation (\ref{taumin}) shows that while all states $\ket{\psi_0}$ eventually become maximally entangled, not all of them become separable along their evolution.

The times $\tilde t_{\min}$ are solutions of the equation $\Delta(\tilde t_{\min})=\Delta_{\max}$, namely 
\begin{equation}
\label{cond_tmin}
(1-2 \, b) \sin 2 \phi\, \cos \pi\tilde t_{\min}\,   = -\cos 2 \phi\,\sin \pi\tilde t_{\min}. 
\end{equation}
Comparison with Eq. (\ref{cond_tmax}) gives for the solutions of (\ref{cond_tmin})
\beq\label{rel_tmaxtmin}
\tilde{t}_{\min}=\tilde{t}_{\max}\pm \frac{1}{2}.
\eeq

\subsection{Entanglement rate}

The maximal loss of entanglement experienced during the evolution is given by
\begin{eqnarray}\label{eq:loss-ent}
\Delta \tau &=&\tau_{\max}-\tau_{\min}\nonumber\\
&=&1-4b(1-b)\tau(0).
\end{eqnarray}
and such amount of entanglement is lost in a time interval given by $\Delta\tilde{t}=1/2$, in line with the relation (\ref{rel_tmaxtmin}). 

\begin{figure}[ht]
    \centering
            \includegraphics[width=.6\linewidth]{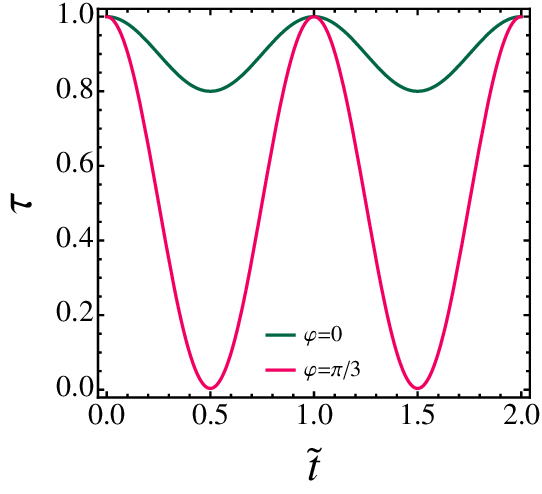}
    \caption{Evolution of the 3-tangle for two initial states $\ket{\psi_0(\phi,\varphi)}$ that are maximally entangled, with $\varphi=0$ (green curve) and $\varphi=\pi/3$ (pink curve) (in both cases we fix $r=2$). The latter exhibits the maximal loss of 3-tangle within a period, therefore has maximal entanglement rate for all $\tilde t\neq \tilde t_{\min,\max}$.}
    \label{fig:tau-loss}
\end{figure}

For fixed Hamiltonian parameters, the period of $\tau(t)$ (Eq. (\ref{tiempoT})) is the same regardless of the initial conditions, yet the entanglement loss depends on the initial state. 
This means that there are states whose 3-tangle evolves more rapidly than others, as can be seen in 
Fig. \ref{fig:tau-loss}, which shows $\tau$ vs $\tilde{t}$ for maximally entangled initial states ($\phi=\pi/4$), with relative phase $\varphi=0$ (green curve) and $\varphi=\pi/3$ (pink curve). 
We clearly see that the loss of entanglement in an interval $[0,\tilde t]$ with $\tilde t<1$ differs significantly between the two states. 
In particular, the state corresponding to the pink curve always has an entanglement velocity greater than the state represented in green, except for the times in which the rate vanishes (maxima and minima of $\tau$). 
This means that the greater the entanglement loss, the greater the maximal entanglement rate. 
From Eq. (\ref{eq:loss-ent}) we thus conclude that the states with maximal entanglement rate will be thus for which $b=0,1$, and in turn these states will eventually become separable, having $\tau_{\min}=0$, as follows from Eq. (\ref{taumin}).  

The above observations can be formally stated by analyzing the entanglement rate 
\beq\label{GammaDef}
\Gamma(t)=\frac{d\tau}{dt}.
\eeq
Resorting to Eqs. (\ref{tau_adim}) and (\ref{ttilde}) we find 
\beq\label{Gamma}
\Gamma(\tilde{t})=T\,\Gamma(t)=A\sin2\pi\tilde{t}+B\cos2\pi\tilde{t},
\eeq
with
\begin{subequations}\label{Gammabis}
\begin{eqnarray}
A
&=&\pi
\Big[\cos 4\phi+4b(1-b)\sin^2 2\phi\Big],\\
B
&=&\pi(1-2b)\sin 4\phi.
\end{eqnarray}
\end{subequations}
Equation (\ref{Gamma}) can be rewritten as 
\beq\label{Gamma simple}
\Gamma(\tilde{t})=\sigma \,C \sin \,(2\pi\tilde{t}+\eta)
\eeq
with $C=\sqrt{A^2+B^2}$, $\eta=\arctan(B/A)$, and 
\begin{eqnarray}\label{sigma}
\sigma=\begin{cases}
			+\,\textrm{sign}(B) & \textrm{if }\,\,\eta\in(0,\pi),\\
            -\,\textrm{sign}(B) & \textrm{if }\,\,\eta \in(\pi,2\pi).
		 \end{cases}
\end{eqnarray}
Direct calculation resorting to (\ref{Gammabis}) gives
\beq\label{C2}
C=\pi\sqrt{
1+4b(1-b)D}
\eeq
where
\beq\label{CBis}
D=2\tau(0)\cos{4\phi}+4b(1-b)\tau^2(0)-\sin^2{4\phi}.
\eeq
Recalling that $\tau(0)=\sin^2 2\phi$, it can be verified that $-1\leq 4b(1-b)D(\phi,b)\leq 0$, whence $-\pi\leq \Gamma(\tilde{t})\leq\pi$. 
\\
\paragraph{Maximal entanglement rate.-}
According to the above results, the maximum value of $\Gamma$ is attained when the following two conditions are met: i) $\sigma=+$, and ii) $4b(1-b)D=0$. 
For fixed (yet arbitrary) $\phi$ the latter condition implies that $b$ should be either $0$ or $1$.

For $b=0$ we get $B=\pi\sin 4\phi$ and $A=\pi\cos 4\phi$, so $\eta=4\phi$. 
Then, according to (\ref{sigma}), the condition $\sigma=+$ will be met provided $\phi\in(0,\pi/4)$. In other words, 
{\bf $\boldsymbol{b=0}$ maximizes $\boldsymbol{\Gamma}$ whenever $\boldsymbol{\phi\in(0,\pi/4)$}}.

For $b=1$ we have $B=-\pi\sin 4\phi$ and $A=\pi\cos 4\phi$, thus
$\eta=2\pi-4\phi$. 
Again, resorting to (\ref{sigma}), the condition $\sigma=+$ holds  provided $\phi\in(\pi/4,\pi/2)$, and we conclude that 
{\bf $\boldsymbol{b=1}$ maximizes $\boldsymbol{\Gamma}$ whenever $\boldsymbol{\phi\in(\pi/4,\pi/2)$}}.

From the expression (\ref{eqn:10}) we find that the condition $b=0$ amounts to
\begin{equation} \label{b=0}
 \frac{r \, \sin \varphi +  \cos \varphi}{ \sqrt{r^{2}+1} } =1.
\end{equation}
For fixed $r$, the solutions of this equation give the \emph{optimal} phases, $\varphi^{\textrm{I}}_\textrm{op}$, that maximize $\Gamma$ whenever $\phi\in(0,\pi/4]$, namely
\begin{equation}
\label{varphioptI}
\varphi^{\textrm{I}}_\textrm{op}=\arctan \,r+2n\pi, \quad \varphi^{\textrm{I}}_\textrm{op}\in\Big(\!\!-\frac{\pi}{2},\frac{\pi}{2}\Big),
\end{equation}
with $n$ an integer, and provided $r\neq0$. 
In particular, $\varphi^{\textrm{I}}_\textrm{op}\in(-\pi/2,0)$ whenever $r<0$, and 
$\varphi^{\textrm{I}}_\textrm{op}\in(0,\pi/2)$ for $r>0$.

On the other hand, the condition 
$b=1$ implies that $r$ and $\varphi$ comply with
\begin{equation} 
\label{b=1}
 \frac{r \, \sin \varphi +  \cos \varphi}{ \sqrt{r^{2}+1} } =-1.
\end{equation}
Solving for $\varphi$ with $r$ fixed, we obtain the optimal phases, $\varphi^{\textrm{II}}_\textrm{op}$, valid for $\phi\in(\pi/4,\pi/2]$:
\begin{equation}
\label{varphioptII}
    \varphi^{\textrm{II}}_\textrm{op}=\arctan\,r+\pi(2n+1),\quad \varphi^{\textrm{II}}_\textrm{op}\in\Big(\frac{\pi}{2},\frac{3\pi}{2}\Big),
\end{equation}
with $n$ an integer, and provided $r\neq0$. 
In particular, $\varphi^{\textrm{II}}_\textrm{op}\in(\pi/2,\pi)$ whenever $r<0$, and 
$\varphi^{\textrm{II}}_\textrm{op}\in(\pi,3\pi/2)$ for $r>0$.

Interestingly, if $\varphi$ coincides with either one of the optimal phases, the relative phase of the evolving state (\ref{psi_equiv}) will remain constant during the evolution, that is, $\varphi(t)=\varphi_{\textrm{opt}}$.
This can be verified by direct substitution of the expressions for $\varphi^{\textrm{I,II}}_{\textrm{opt}}$ in Eqs. (\ref{eqn:6}) and (\ref{alpha2}). 

\begin{figure}[ht]
    \centering
     \includegraphics[width=0.8\columnwidth]{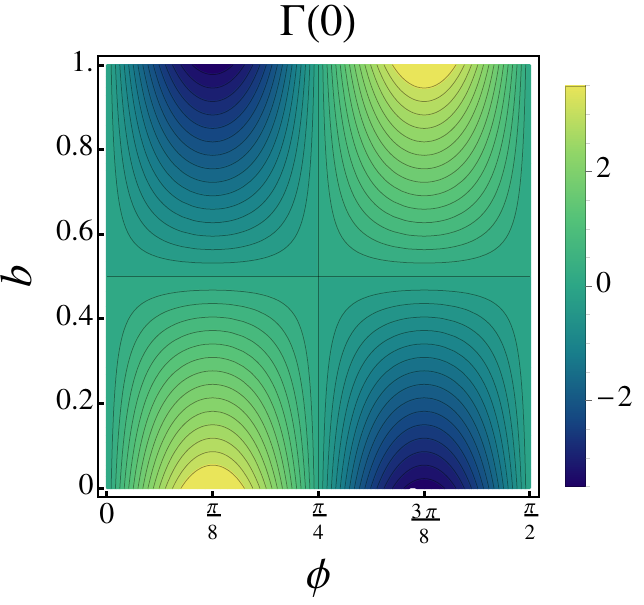}
    \caption{Initial entanglement rate $\Gamma_0$ (Eq. (\ref{Gamma00})), as a function of $b$ and $\phi$. For fixed $\phi$, $\Gamma_0$ attains its maximal value at $b=0$ provided $\phi\in (0,\pi/4)$, and at $b=1$ for $\phi\in (\pi/4, \pi/2)$.}
    \label{fig:Gamma0Bis}
\end{figure}

As follows from Eq. (\ref{Gamma}) the \emph{initial} entanglement rate $\Gamma(0)=\Gamma_0$ is 
\beq\label{Gamma00}
\Gamma_0=B=\pi(1-2b)\sin 4\phi.
\eeq
For $\phi\in(0,\pi/4]$ the sign of $\Gamma_0$ is the sign of $1-2b$, so the entanglement initially increases whenever $b<1/2$, and decreases for $b>1/2$. 
According to the previous results, in this case the optimal value of $b$ is $0$ (so the optimal phase is given by $\varphi^{\textrm I}_{\textrm{op}}$). As for $\phi$, its optimal value (the one that maximizes $\Gamma_0$ in the interval $\phi\in(0,\pi/4]$) reads $\phi^{\textrm{I}}_\textrm{op}=\pi/8$. Analogously, for $\phi\in(\pi/4,\pi/2]$ 
the sign of $\Gamma_0$ is the sign of $2b-1$, so the entanglement initially increases whenever $b>1/2$, and decreases for $b<1/2$. 
For fixed $\phi$ in this interval, $\Gamma_0$ reaches it maximum value when the relative phase is $\varphi^{\textrm{II}}_\textrm{op}$,
and in this case, the optimal value for $\phi$ is given by $\phi^{\textrm{II}}_\textrm{op}=3\pi/8$.
Figure \ref{fig:Gamma0Bis} shows  $\Gamma_0$ as a function of $b$ and $\phi$. It clearly verifies that for
$\phi\in (0,\pi/4]$, the maximum is found at  
$\phi=\phi^{\textrm{I}}_\textrm{op}=\pi/8$ and $\varphi=\varphi^{\textrm{I}}_\textrm{op}$ ($b=0$), whereas for $\phi\in (\pi/4,\pi/2]$ it is attained at 
$\phi=\phi^{\textrm{II}}_\textrm{op}=3\pi/8$ and $\varphi=\varphi^{\textrm{II}}_\textrm{op}$ ($b=1$).
%

Finally, the sign of $\Gamma_0$ for different ranges of $\phi$ and the conditions that must be fulfilled by $r$ and $\varphi$ are those indicated in Table \ref{tabla1}.
\begin{table}[ht]
        \centering
        \begin{tabular}{c||c||c}
                & $\phi\in[0,\pi/4]$ & $\phi\in[\pi/4,\pi/2]$\\    
        \hline
        \hline
            $0\leq r\sin\varphi+\cos\varphi\leq\sqrt{r^2+1}$ &+ & -- \\
            $-\sqrt{r^2+1}\leq r\sin\varphi+\cos\varphi\leq 0$ &-- &  +\\
         
        \end{tabular}
        \caption{Sign of $\Gamma_0$ depending on the Hamiltonian and the initial state parameters.}
        \label{tabla1}
    \end{table}  

\section{Optimization of the entanglement rate via single-qubit operations}\label{opti}

\subsection{Optimization of $\Gamma$ via a  rotation}\label{sec:opt-gamma1}

Consider the initial state $\ket{\psi_0}=\ket{\psi_0(\phi,\varphi)}$, given by Eq. (\ref{eqn:1}).
In general, the corresponding $\Gamma_0$ will not be maximal, not even will it be positive, so one would have to wait until $\tilde t=\tilde t_{\max}$ to have a maximally entangled state. 
However, the time required to reach such state can be shortened by applying a rotation that transforms the state $\ket{\psi_0}$ into  
\begin{eqnarray}\label{rotado}
\ket{\psi_0^\textrm{op}}&=& R_\textrm{op}|\psi_0(\phi,\varphi)\rangle=\ket{\psi_0(\phi,\varphi_{\textrm{op}})}\\
\nonumber
&=&\sin\phi|000\rangle+e^{i\varphi_{\textrm{op}}}\cos\phi|111\rangle,
\end{eqnarray}
with $\varphi_{\textrm{op}}$ either $\varphi^{\textrm{I}}_{\textrm{op}}$ or $\varphi^{\textrm{II}}_{\textrm{op}}$ , depending on whether $\phi$ is in $(0,\pi/4)$, or in $(\pi/4,\pi/2)$. 
Since $R_\textrm{op}$ can be decomposed as
\beq \label{rot}
R_\textrm{op}=\mathbb I\otimes \mathbb I\otimes U_\textrm{op},
\eeq
with $U_\textrm{op}$ represented by the unitary matrix
\beq U_\textrm{op}(\varphi_{\textrm{op}}-\varphi)=\begin{pmatrix}
1 & 0 \\
0 & e^{i(\varphi_{\textrm{op}}-\varphi)} 
\end{pmatrix},
\eeq
the mapping $\ket{\psi_0}\rightarrow R_\textrm{op}\,|\psi_0\rangle$ is a local unitary transformation that leaves the entanglement of the state unaffected. 
Further, by design, it enhances the entanglement rate, thus leaving us with a state (\ref{rotado}) that has the maximal $\Gamma_0$ given the value of the initial parameter $\phi$. 

\begin{figure}
    \centering
\includegraphics[width=0.8\columnwidth]
{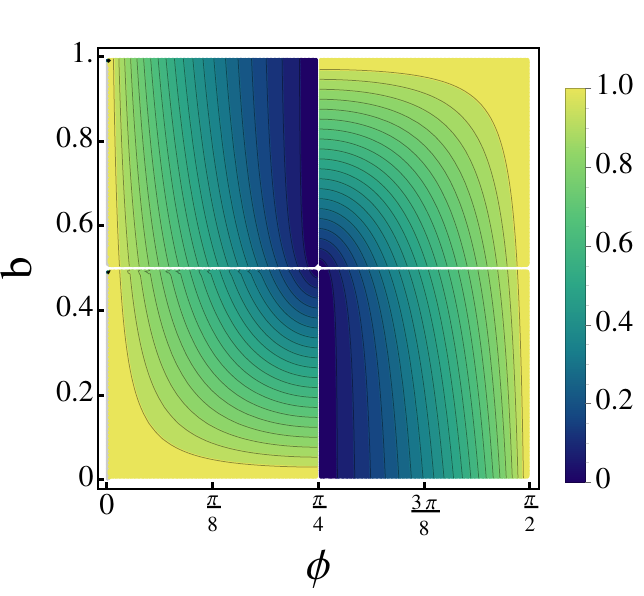}
    \caption{Ratio $\tilde t^{\prime}_{\max}/\tilde t_{\max}$ between the time it takes $\ket{\psi^{\textrm{op}}_0}$ and $\ket{\psi_0}$ to evolve towards a maximally entangled state. The transformation $\ket{\psi_0}\rightarrow R_\textrm{op}\,|\psi_0\rangle$ reduces the time needed to reach $\tau=1$, and the higher reduction occurs in the quadrants in which $\Gamma_0<0$. }
    \label{tiempored}
\end{figure}

The time the new initial state $\ket{\psi_0^\textrm{op}}$ takes to become a maximally entangled state is now
\beq
\tilde t^{\prime}_{\max}=\begin{cases}
			\tilde t_{\max}|_{b=0}=\frac{1}{2}-\frac{2\phi}{\pi}<\frac{1}{2}& \textrm{if }\,\,\phi\in(0,\frac{\pi}{4}),\\
            \tilde t_{\max}|_{b=1}=\frac{2\phi}{\pi}-\frac{1}{2}<\frac{1}{2} & \textrm{if }\,\,\phi \in(\frac{\pi}{4},\frac{\pi}{2}).
		 \end{cases}
\eeq
In order to compare $\tilde t^{\prime}_{\max}$ with the time it would take $\ket{\psi_0}$ to attain the maximal 3-tangle  ($\tilde t_{\max}$ given by Eq. (\ref{tmax})), we focus on the ratio $\tilde t^{\prime}_{\max}/\tilde t_{\max}$, which gives the factor by which the time is reduced when the rotation $R_{\textrm{op}}$ is performed. 
Figure \ref{tiempored} shows the plot of $\tilde t^{\prime}_{\max}/\tilde t_{\max}$ as a function of $\phi$ and $b$. As expected, the time reduction increases ($\tilde t^{\prime}_{\max}/\tilde t_{\max}$ attains lower values) in the quadrants where $\Gamma_0$ is negative (c.f Fig. \ref{fig:Gamma0Bis}).

\begin{figure}
    \centering
    \includegraphics[width=0.45\columnwidth]{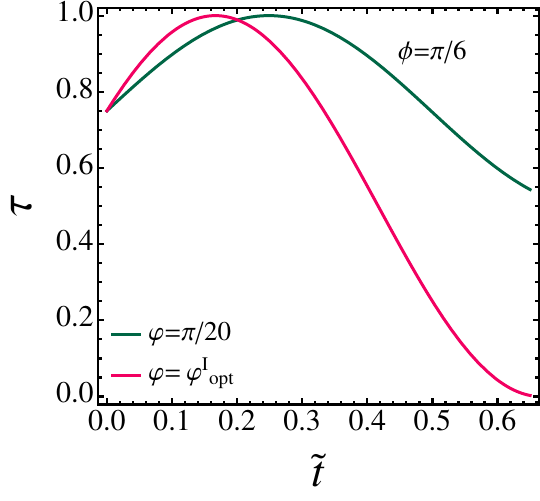}
    \includegraphics[width=0.45\columnwidth]{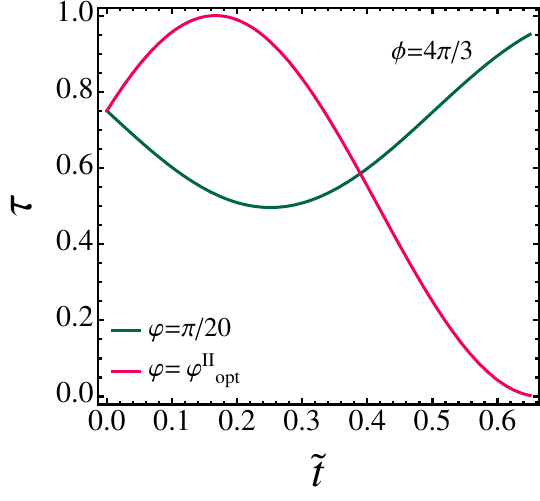}
    \caption{Evolution of the 3-tangle for an initial state $\ket{\psi_0}$ (green curve) and for the rotated state $\ket{\psi^{\textrm{op}}_0}$ (pink curve) considering $\Gamma_0>0$ (left panel), and $\Gamma_0<0$ (right panel). In both cases we fix $r=2$. The reduction in the time needed to attain $\tau=1$ is manifest. }
    \label{fig:opt-varphi}
\end{figure}

Figure~\ref{fig:opt-varphi} depicts the improvement in the time needed to reach the maximum 3-tangle when shifting the relative phase to its optimal value, by showing the dynamics of the 3-tangle. 
The green curve corresponds to the evolution of $\tau$ for the state $\ket{\psi_0}$, whereas the pink one depicts the evolution of $\tau$ for $\ket{\psi_0^\textrm{op}}$. 
In the left panel the corresponding initial state has $\Gamma_0>0$, and the optimization is slightly significative, yet in the right panel the initial state has $\Gamma_0<0$ and the time required to achieve maximal entanglement its reduced more drastically. 

The fact that $\ket{\psi_0^\textrm{op}}$ has the optimal phase, guarantees that it is the state that attains $\tau=1$ \emph{faster} than any other state with the same initial 3-tangle.
\subsection{Optimization of $\Gamma$ involving a qubit flip}\label{sec:flop-vs-phase}
Given that the states considered have two parameters, $\phi$ and $\varphi$, a fair and interesting question is whether an optimization of the entanglement rate can also be made by operating on the population's parameter, $\phi$. 
In general, a change in this parameter would modify the coefficients of the states $\ket{000}$ and $\ket{111}$, which in turn would induce a change in the amount of 3-tangle. 
Consequently, the entanglement rate cannot in general be maximized by changing the value of $\phi$ in a local fashion.
However, from Eq. (\ref{tau}) we see that for fixed $\tau$ there are two (positive) possible values of $p$, namely
\begin{equation}\label{sols}
    p_\pm=\frac{1\pm\sqrt{1-\tau}}{2},
\end{equation}
where the relation $p_++p_-=1$ can be readily seen. 

Let us consider the initial states $|\psi^\pm(0)\rangle=\sqrt{p_\pm}\,|000\rangle+e^{i\varphi}\sqrt{1-p_\pm}\,|111\rangle$. Applying 
a (local) spin flip operation $ U_f=\sigma_x\otimes\sigma_x\otimes\sigma_x$ gives
\begin{eqnarray}\label{eq:op-flip} U_f|\psi^\pm(0)\rangle&=& U_f(\sqrt{p_\pm}\,|000\rangle+e^{i\varphi}\sqrt{1-p_\pm}\,|111\rangle)\nonumber\\
    &=&\sqrt{p_\pm}\,|111\rangle+e^{i\varphi}\sqrt{1-p_\pm}\,|000\rangle \nonumber\\
    &=&\sqrt{1-p_\mp}\,|111\rangle+e^{i\varphi}\sqrt{p_\mp}\,|000\rangle\nonumber\\
    &\sim & \sqrt{p_\mp}\,|000\rangle+e^{-i\varphi}\sqrt{1-p_\mp}\,|111\rangle,
\end{eqnarray}
where in the last line $\sim$ indicates equivalence of the states up to a global phase. 
That the flip does not affect the amount of $\tau$ follows straightforward from the fact that $U_f$ is a local operation. Further, it amounts to exchange
\beq\label{flip}
p\leftrightarrow (1-p),\quad\textrm{and}\quad \varphi \leftrightarrow -\varphi,
\eeq
which from Eq. (\ref{tau}) clearly leaves the 3-tangle unaffected.

The transformation $p\leftrightarrow (1-p)$, together with the identification $\sqrt{p}=\sin\phi$, is equivalent to perform a reflection along the direction $\pi/4$, so the exchange in the populations corresponds to
\beq\label{transf_phi}
\phi\leftrightarrow (\pi/2)-\phi,
\eeq
and the effect of the spin-flip on a generic initial state $\ket{\psi_0}=\ket{\psi_0(\phi,\varphi)}$ thus reads
\beq\label{fliprule}
\ket{\psi^{\textrm{f}}_0}=U_f\ket{\psi_0(\phi,\varphi)}=\ket{\psi_0(\tfrac{\pi}{2}-\phi,-\varphi)}.
\eeq

The transformation (\ref{transf_phi}) maps the interval $[0,\pi/4]$ into $[\pi/4,\pi/2]$, and exchanges $\Gamma_0\leftrightarrow -\Gamma_0$, as can be directly verified substituting Eq. (\ref{transf_phi}) into (\ref{Gamma00}). 
The second transformation in (\ref{flip}) may, however, also affect $\Gamma_0$, via its dependence on $B(\varphi)=1-2b(\varphi)$.
Now, the signs of $B(\varphi)$ and $B(-\varphi)$ coincide if and only if their product is positive, which occurs (following Eq. (\ref{eqn:10}) and ruling out the case $\varphi=0$), whenever 
\beq\label{cond_signo}
\cot^2\varphi>r^2.
\eeq
Thus, when the condition (\ref{cond_signo}) is met, the exchange $\varphi\leftrightarrow -\varphi$ does not alter the sign of $\Gamma_0$, and the net effect of the flip operation is  
\beq
\textrm{sgn}\,\Gamma_0\leftrightarrow -\textrm{sgn}\,\Gamma_0.
\eeq
In this way, and provided Eq. (\ref{cond_signo}) holds, a single spin flip transforms an initial state with $\Gamma_0<0$ into one with positive entanglement rate, thus reducing the time needed to attain a state with maximal $\tau$. 
If $\varphi=0$ the flip transformation would correspond to the mapping $|\psi^\pm(0)\rangle \leftrightarrow |\psi^\mp(0)\rangle$, and would suffice to guarantee that  $\Gamma_0$ becomes $-\Gamma_0$.
%

\begin{figure}
    \centering
    \includegraphics[width=.45\linewidth]{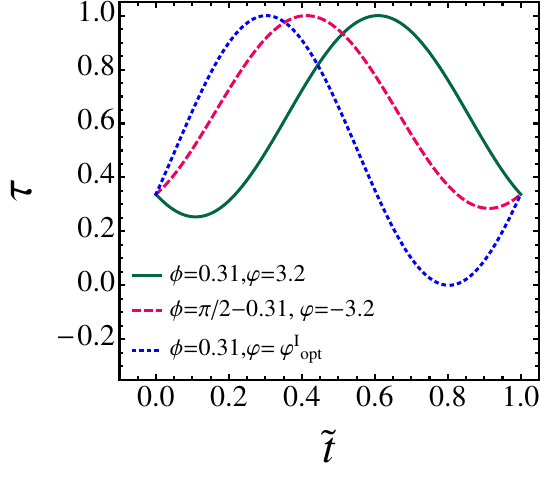}
\includegraphics[width=.45\linewidth]{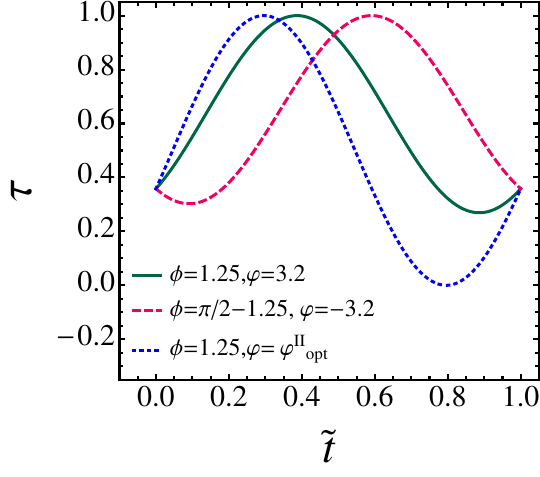}
    \caption{Evolution of the 3-tangle for the initial states $\ket{\psi_0}$ (green solid line), $\ket{\psi^{\textrm{f}}_0}$ (pink dashed line), and $\ket{\psi^{\textrm{op}}_0}$ (blue dotted line), considering $\Gamma_0<0$ (left panel) and $\Gamma_0>0$ (right panel). In all cases we fix $r=2$, and the parameters are such that the condition (\ref{cond_signo}) is satisfied. }
    \label{fig:signo-cond}
\end{figure}

Notice, however, that this procedure does not involve a maximization of the entanglement rate, so in general the improvement in the speed towards $\tau=1$ will be less than the (maximal) improvement induced by the previous protocol. 
A comparison of the two strategies can be seen in Fig. \ref{fig:signo-cond}. There, an initial state $\ket{\psi_0}$ whose parameters comply with (\ref{cond_signo}) is considered, and the evolution of $\tau$ is shown for the cases in which: the initial state evolves solely under the action of the Hamiltonian (green solid curve); the flip operation (\ref{fliprule}) is performed on the state (pink dashed curve); the rotation (\ref{rot}) that optimizes the relative phase is applied (blue dotted curve). 
The left panel corresponds to an initial state with  $\Gamma_0<0$, whereas the right one to a state with $\Gamma_0>0$.     
As expected, the optimization procedure by means of the rotation (\ref{rotado}) always results in a shorter time to reach $\tau=1$. Further, it can be seen that the spin-flip reduces the time to attain the maximal entanglement provided  $\Gamma_0<0$.


We have seen that even if (\ref{cond_signo}) holds, the flip operation will not always be useful, specifically when $\Gamma_0>0$. 
Moreover, if $\ket{\psi_0(\phi,\varphi)}$ is such that the condition  (\ref{cond_signo}) is not met, the flip operation will not invert the sign of the (possibly negative) initial entanglement rate. 
However, $U_f$ can still reduce the time needed to reach $\tau=1$ provided $\Gamma_0( U_f|\psi_0\rangle)>\Gamma_0(|\psi_0\rangle)$, i.e., provided
\begin{eqnarray}
\Gamma_0(\phi,\varphi)<\Gamma_0\Big(\frac{\pi}{2}-\phi,-\varphi\Big).
\end{eqnarray}
Resorting to Eq. (\ref{Gamma00}), this leads to the condition 
\begin{equation}\label{cond_signo_2}
    \cos\varphi\,\sin 4\phi<0,
\end{equation}
which implies that the flip operation is useful also (despite (\ref{cond_signo}) is not met), in the following cases
\begin{eqnarray}\label{cond2}
\varphi&\in& \Big(\frac{\pi}{2}, \frac{3\pi}{2}\Big),\;\phi\in \Big(0, \frac{\pi}{4}\Big)\Rightarrow \Gamma_0>0;\\
\varphi&\in& \Big(\frac{3\pi}{2}, \frac{\pi}{2}\Big),\;\phi\in \Big(\frac{\pi}{4}, \frac{\pi}{2}\Big)\Rightarrow \Gamma_0<0.\nonumber
\end{eqnarray}

\begin{figure}[ht]
    \centering
    \includegraphics[width=0.8\linewidth]{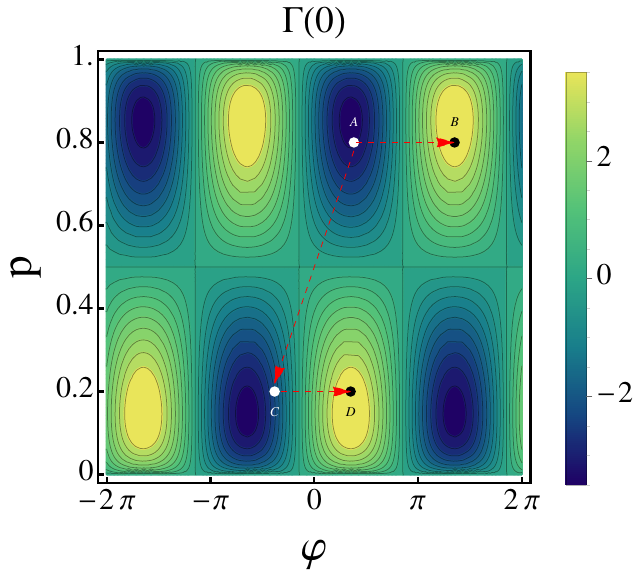}
    \caption{Two possible paths in the space $(\varphi,p)$, namely $A\rightarrow B$ and $A\rightarrow C\rightarrow D$, to maximize the initial entanglement rate $\Gamma_0$ (color scale). The point $A$ represents the initial state $\ket{\psi_0}$, $B$ represents the rotated state $R_{\textrm{op}}\ket{\psi_0}$, $C$ corresponds to $U_{f}\ket{\psi_0}$, and $D$ to $R_{\textrm{op}}(U_{f}\ket{\psi_0})$. Both final states, $B$ and $D$, have the same $\tau$ and entanglement rate. The Hamiltonian parameter was fixed to $r=2$. }
    \label{fig:path_gama}
\end{figure}

The flip operation can be improved by performing an additional rotation that shifts the phase $-\varphi$ in the last line of (\ref{eq:op-flip}) to some appropriate value. 
Clearly, the best possible choice is to perform a rotation (\ref{rot}) that brings the relative phase to its optimal value. 
Such flip-plus-rotation procedure is equivalent to the single-rotation operation that maximizes the entanglement rate, thus giving two paths to optimize  $\Gamma_0$, depicted in 
Figure~\ref{fig:path_gama}. 
%

Let us first consider the initial state $\ket{\psi_0}$, represented by the point $A$ and determined by $p_A=0.8$ and $\varphi_A=1.2$. A phase optimization via $R_{\textrm{op}}$ leads to the optimal state $\ket{\psi^{\textrm{op}}_0}$, represented by the point $B$, whose parameters are $p_B=0.8$ and $\varphi_B=4.248$. This gives the first path, $A\rightarrow B$, that maximizes $\Gamma_0$.  
A second way to optimize the entanglement rate  involves performing the flip operation on the state $\ket{\psi_0}$, which leads to the intermediate state $\ket{\psi^{\textrm{f}}_0}$ ---corresponding to $p_C=0.2$, $\varphi_C=-1.2$ and to the point $C$---, and then optimizing the phase by rotating $\ket{\psi^{\textrm{f}}_0}$, which leads to the final state represent by the point $D$, and determined by $p_D=0.2$ and $\varphi_D=1.107$. This gives the second path $A\rightarrow C\rightarrow D$.
By construction, both final states, corresponding to $B$ and $D$, have the same amount of $\tau$, and the same entanglement rate,  $\Gamma(\phi_B,\varphi_B)=\Gamma(\phi_D,\varphi_D)$, as can be appreciated in Figure \ref{fig:path_gama}. 
The fact that $\Gamma(\phi_B,\varphi_B)=\Gamma(\phi_D,\varphi_D)$ can  be easily proven in general for an arbitrary initial state $|\psi_0(\phi,\varphi)\rangle$ by using Eq. (\ref{transf_phi}) and the relation between the two possible optimal values of $\varphi$ (Eqs. (\ref{varphioptI}) and (\ref{varphioptII})). A simple calculation leads to $\Gamma_0(\phi,\varphi_{\textrm{op}}^{\textrm{I,II}})=\Gamma_0(\pi/2-\phi,\varphi_{\textrm{op}}^{\textrm{II,I}})$, states that correspond to $B$ and $D$ in the previous example.

The results shows that for each initial state $\ket{\psi_0}$, there are two different (yet equally entangled) optimal states ($R_{\textrm{op}}\ket{\psi_0}$ and $R_{\textrm{op}}[U_f\ket{\psi_0}]$), with improved, maximal, entanglement rate. The analysis also suggests that it is more convenient to optimize the entanglement rate directly shifting the phase to its optimal value, instead of flipping the qubits and thereafter adjusting the phase, which requires more time invested in the process. 

Despite its cons, the flip operation has the advantage of involving a \emph{fixed, universal} transformation $ U_f=\sigma_x\otimes \sigma_x\otimes \sigma_x$ that serves for improving the entanglement rate of a wide class of initial states, namely those  that comply with  (\ref{cond_signo}) (with $\Gamma_0<0$) or with (\ref{cond2}), without the need of implementing a calibrated rotation. 

\section{Preventing the entanglement from decaying via single-qubit operations}

We now present different protocols aimed at maintaining the 3-tangle above some predetermined threshold value $\tau^*$ for all times. They are based on the application of the local operations described above, as a means to 
counteract the loss of 3-tangle induced by the Hamiltonian evolution. 

\begin{figure} 
\centering
    \includegraphics[width=.6\linewidth]{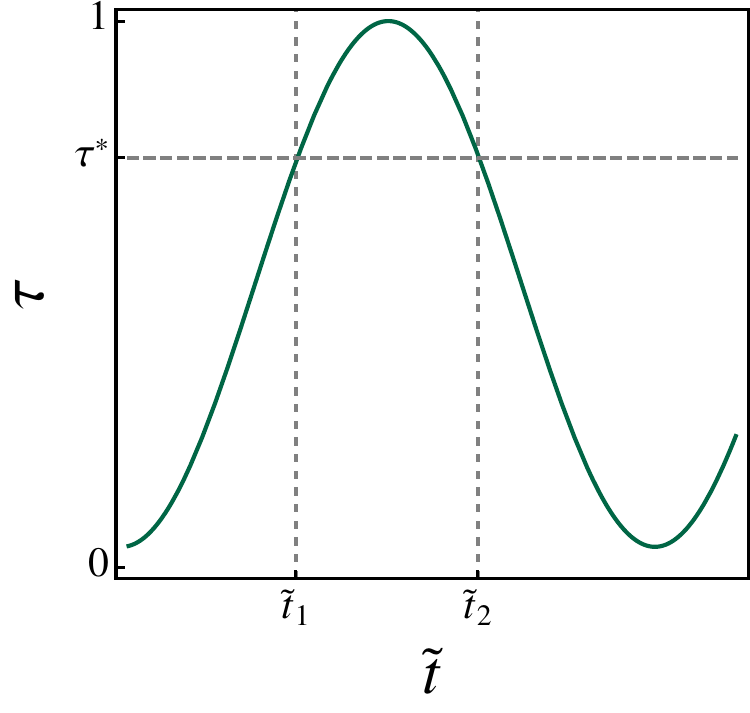}
    \caption{Solutions $\tilde t_{1,2}$ of Eq. (\ref{eq:sol-tau-umbral}) such that for $\tilde t\in [\tilde t_{1},\tilde t_{2}]$ the 3-tangle does not drop below a threshold value $\tau^*$.}
    \label{fig:thresh}
\end{figure}

Firstly, in order to determine the length of the time steps between local operations,  it is 
helpful to write the 3-tangle in the form 
\begin{equation}
    \tau(\tilde t)=1-A^2\cos^2(\pi\tilde t + \chi).
\end{equation}
Comparison of this expression with Eq. (\ref{tau_adim}) gives $A^2=1-\tau_{\min}$, and $\tan \chi=(1-2b)\tan 2\phi$. 
Therefore, the times $\tilde t^*$ at which $\tau$ attains the threshold value $\tau^*$ are the solutions of the equation
\begin{equation}\label{eq:sol-tau-umbral}
    |\cos(\pi\tilde t^*+\chi)|=\sqrt{\frac{1-\tau^*}{1-\tau_{\min}}}.
\end{equation}
As depicted in Fig.~\ref{fig:thresh}, this equation has two solutions within the first period, denoted as $\tilde{t}_1$ and $\tilde{t}_2>\tilde{t}_1$. 
Assuming that $\Gamma_0>0$, at $\tilde{t}_1$ the entanglement rate is positive, so for  $\tilde{t}\in[\tilde{t}_1,\tilde{t}_2]$ we have $\tau(\tilde{t})\geq \tau^*$. 
At $\tilde{t}_2$ the entanglement rate is negative and immediately afterwards the 3-tangle will drop below the threshold value. 
Therefore, the maximal amount of time that we can let the state to evolve after $\tilde{t}_1$ in order to prevent $\tau$ from dropping below $\tau^*$ is $\delta\tilde{t}=\tilde{t}_2-\tilde{t}_1$. 
At this point, a local operation must be applied to reverse the loss of entanglement, as we will discuss in the next subsections.

\subsection{Keeping $\tau$ above a threshold value via $\sigma_z$}\label{gelatina}

As seen in Section~\ref{sec:opt-gamma1}, for any initial state $\ket{\psi_0}$ shifting the relative phase to its optimal value $\varphi_{\textrm{op}}$ will maximize its (initial) entanglement rate. 
The rotation (\ref{rot}) that produces this shift is thus the first operation applied in the protocol. 

After the transformation $\ket{\psi_0}\rightarrow \ket{\psi^{\textrm{op}}_0}$ we have a state with $\varphi=\varphi_{\textrm{op}}$ and increasing entanglement, as in Fig.~\ref{fig:thresh}.
As $\ket{\psi^{\textrm{op}}_0}$ evolves, its relative phase will keep constant (recall the discussion below Eq. (\ref{varphioptII})), while its 3-tangle tends to its maximum value, reached at $\tilde t=\tilde{t}_{\max}$. 
At that point, the population angle of the state becomes $\phi(\tilde{t}_{\max})=\pi/4$, as corresponds to a maximally entangled state. Afterwards, $\phi(\tilde t)$ will lie in the interval in which the original relative phase $\varphi_{\textrm{op}}$ is not longer the optimal one for the evolving state, which thus starts losing entanglement.
We can then wait until (at most) $\tilde t=\tilde{t}_1+\delta \tilde t$ to apply a second rotation that fixes the relative phase to its new corresponding optimal value, which is either $\varphi^{\textrm{I}}_\textrm{op}$ or $\varphi^{\textrm{II}}_\textrm{op}$, depending on whether the original phase   was  $\varphi^{\textrm{II}}_\textrm{op}$ or $\varphi^{\textrm{I}}_\textrm{op}$, respectively. 
The second step in the protocol thus exchanges $\varphi^{\textrm{I}}_\textrm{op}\leftrightarrow\varphi^{\textrm{II}}_\textrm{op}$. 
As follows from Eqs. (\ref{varphioptI}) and (\ref{varphioptII}),  $\varphi^{\textrm{II}}_\textrm{op}=\varphi^{\textrm{I}}_\textrm{op}+\pi$, whence the required rotation is one of the form (\ref{rot}), with $U_\textrm{op}$ given by
\beq\label{rotz}
U_\textrm{op}= U_\textrm{op}[\pm(\varphi^{\textrm{II}}_\textrm{op}-\varphi^{\textrm{I}}_\textrm{op})]=U_\textrm{op}(\pm\pi)=\sigma_z.
\eeq

Once this rotation is performed, the Hamiltonian evolution is resumed from a new initial state which, by construction, has again a positive entanglement rate and $\tau\geq\tau^*$. 
The 3-tangle increases again, reaches its maximal value and decreases afterwards. 
At a subsequent time $\tilde t=\tilde{t}_1+2\delta \tilde t$ the situation is similar to that at $\tilde t=\tilde t_1+\delta \tilde t$, and an appropriate rotation is again required to prevent $\tau$ from dropping below the threshold value.
Clearly such rotation performs again the exchange $\varphi^{\textrm{I}}_\textrm{op}\leftrightarrow\varphi^{\textrm{II}}_\textrm{op}$, and is implemented via the inverse of (\ref{rotz}), that is, by the Pauli operator $\sigma_z$ again. 

\begin{figure}[ht]
    \centering
    \includegraphics[width=.6\linewidth]{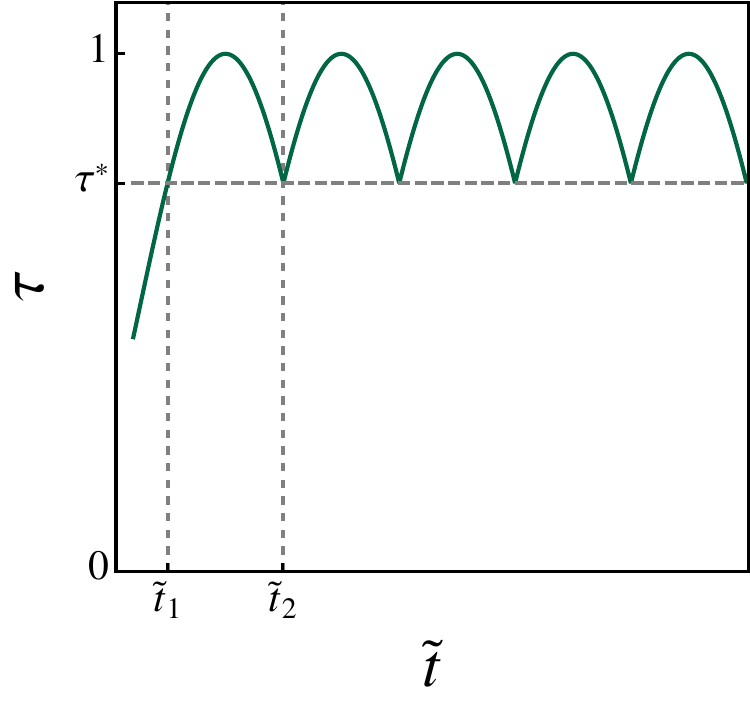}
    \caption{Evolution of the 3-tangle under the protocol (\ref{prot_gelatina}), preventing $\tau(\tilde t)$ to drop below the threshold $\tau^*$.}
    \label{fig:prot-umbral}
\end{figure}

This protocol can be synthesized in the following sequence of operations applied to a single qubit:
\begin{eqnarray}\label{prot_gelatina}
\ket{\psi_0}\xrightarrow{t=0} U_\textrm{op}(\varphi_{\textrm{op}}-\varphi)\xrightarrow{t=t_1+\delta \tilde t}\sigma_z\nonumber\\
\xrightarrow{t=t_1+2\delta \tilde t}\sigma_z\xrightarrow{t=t_1+3\delta \tilde t}\sigma_z\dots,
\end{eqnarray}
where it must be understood that between these operations the evolution is dictated by the Hamiltonian (\ref{Hamiltonian}). The resulting evolution of the 3-tangle is depicted in Fig.~\ref{fig:prot-umbral}.
Apart from the first step in the protocol, which requires a calibrated phase shift of $\varphi_{\textrm{op}}-\varphi$ that accelerates the arrival at a maximally entangled state, the rest of the transformations involved are state-independent Pauli operators. In this way, by passing a single qubit through a fixed gate at time steps of length $\leq\delta \tilde t$ the 3-tangle can be maintained above the desired value $\tau^*$.

\subsection{Reaching a maximally entangled stationary state via a rotation}

As discussed above Eq. (\ref{tauconstant}), the state $\ket{\psi_\textrm{s}}$ characterized by  $\phi_{\textrm{s}}=\pi/4$ and $\varphi_\textrm{s}=\arctan(-1/r)$, is a stationary state with maximal 3-tangle. Approximating an arbitrary initial state $\ket{\psi_0}$ into $\ket{\psi_\textrm{s}}$ will thus prevent any decay in $\tau$, and can be optimally done by means of a pair of suitable phase shifts, as follows. 

Given an initial state $\ket{\psi_0(\phi,\varphi)}$ 
%
\begin{figure}
    \centering
\includegraphics[width=.65\linewidth]{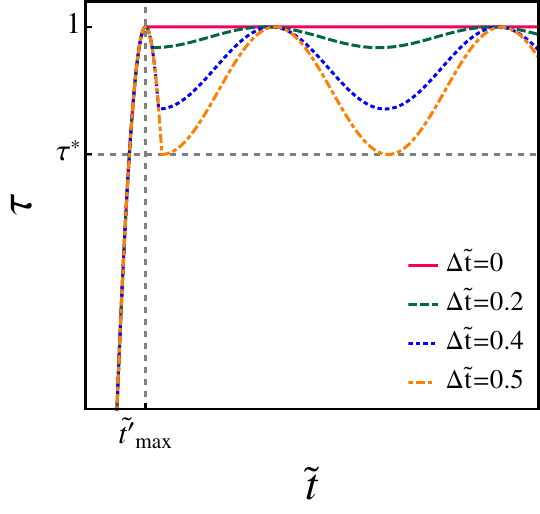}
    \caption{Evolution of the 3-tangle starting from the optimal state $\ket{\psi_0(\phi,\varphi_\textrm{op})}$ with $\phi=0.36, \varphi_\textrm{op}=1.107$, and $r=2$. At $\tilde t=\tilde t^{\prime}_{\max}+\Delta \tilde t$ the rotation $R_{\textrm{s}}$ is applied and the resulting state is let to evolve freely under the action of the Hamiltonian. As $\Delta \tilde t$ increases, the minimum 3-tangle of the evolving state decreases, so by an adequate control of $\Delta \tilde t$ the 3-tangle can be made to oscillate between $1$ and a certain $\bar\tau$ greater than the threshold value $\tau^*$ (represented in the example by the gray dashed line). }
    \label{fig:evol-libre}
\end{figure}
\begin{figure}
    \centering
\includegraphics[width=0.65\linewidth]{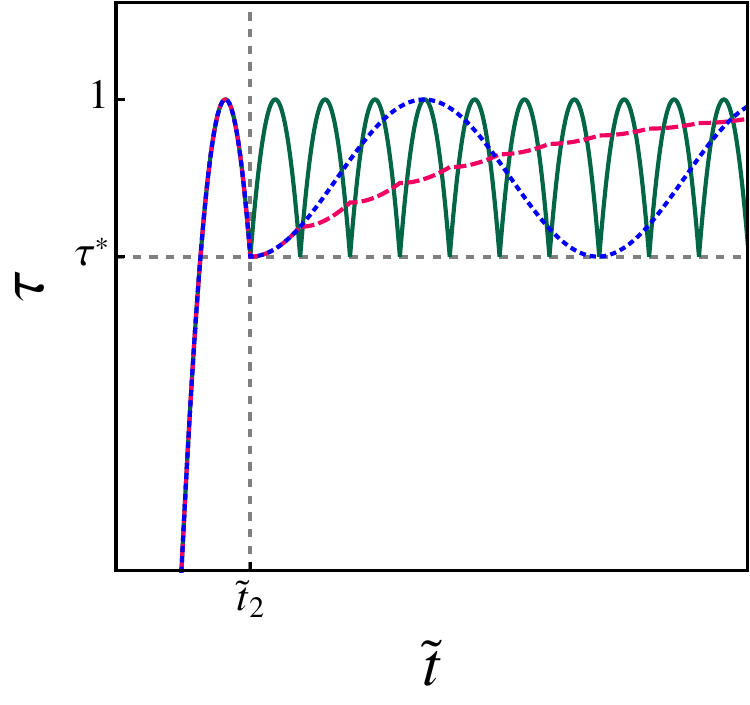}
    \caption{Evolution of the 3-tangle starting from the optimal state $\ket{\psi^{\textrm{op}}_0}$ that evolves with maximal entanglement rate. At  $\tilde t_2=\tilde t_{1}+\delta \tilde t$  three situations are considered: the operator $\sigma_z$ is successively applied, following the protocol (\ref{prot_gelatina}) (solid green curve); 
the rotation $R_{\textrm{s}}$ is periodically applied at intervals of length $\delta\tilde t$ (red dashed curve); 
the rotation  $R_{\textrm{s}}$ is performed only once (blue dotted curve).     }
  \label{fig:optima-magica-libre}
\end{figure}
we apply the rotation (\ref{rotado}), obtaining the state $\ket{\psi_0^\textrm{op}}=\ket{\psi_0(\phi,\varphi_{\textrm{opt}})}$ that optimally evolves towards a state with maximum entanglement, as illustrated at the first stage of the evolution in Figure \ref{fig:evol-libre}. 
 When the maximum entanglement is reached, at $\tilde t=\tilde t^\prime_{\max}$ (meaning that $\phi$ has reached the value $\phi=\pi/4$),  
 we perform a new rotation that shifts the actual phase of $|\psi^\textrm{op}(\tilde t^\prime_{\max})\rangle$ to the value $\varphi_\textrm{s}$, i.e., we apply the 
operator
\beq \label{rots}
R_\textrm{s}=\mathbb I\otimes \mathbb I\otimes U_\textrm{s},
\eeq
with $U_\textrm{s}$ represented by the unitary matrix
\beq 
U_\textrm{s}(\varphi_{\textrm{s}}-\varphi_{\textrm{op}})=\begin{pmatrix}
1 & 0 \\
0 & e^{i(\varphi_{\textrm{s}}-\varphi_{\textrm{op}})} 
\end{pmatrix},
\eeq
and obtain the renewed state $R_\textrm{s}|\psi^\textrm{op}(\tilde t^\prime_{\max})\rangle=\ket{\psi_0(\phi_\textrm{s},\varphi_\textrm{s})}=\ket{\psi_{\textrm{s}}}$. From that point on, the 3-tangle will remain in its highest value $\tau=1$, as depicted in the solid red line in Fig. \ref{fig:evol-libre}. 
  
If, however, there is some delay in the implementation of $R_\textrm{s}$, so that it is applied at $\tilde t=\tilde t^\prime_{\max}+\Delta \tilde t$, the 3-tangle will oscillate between $\tau(\tilde t^\prime_{\max}+\Delta \tilde t)$ and $\tau_\textrm{max}=1$, as can be observed in the 
dashed green, dotted blue, and dot-dashed orange curves, corresponding to different (respectively increasing) values of $\Delta\tilde t$. This oscillating behaviour can be understood as follows. 

As discussed when Eq. (\ref{tauconstant}) was introduced, the phase $\varphi_{\textrm{s}}$ corresponds to $b=1/2$, that is, to $\Gamma_0=0$ (see Eq. (\ref{Gamma00})). 
This means that a state with $\varphi=\varphi_{\textrm{s}}$ either possess maximal ($\tau=1$) or minimal ($\tau=\tau_{\min}$) 3-tangle. The former case occurs provided $\phi=\pi/4$, whereas for any other $\phi$ the condition of vanishing $\Gamma_0$ implies that the state's 3-tangle is a minimum. Therefore, the state
\begin{eqnarray}
R_{\textrm{s}}|\psi^\textrm{op}(\tilde t^\prime_{\max}+\Delta \tilde t)\rangle=\ket{\psi_0(\phi,\varphi_\textrm{s})}
\end{eqnarray}
with $\phi=\phi(\tilde t^\prime_{\max}+\Delta \tilde t)$ and $0<\Delta \tilde t<1$, has a 3-tangle that fixes the minimal $\tau$ in the subsequent Hamiltonian evolution of $\ket{\psi_0(\phi,\varphi_\textrm{s})}$, so that 
\beq
\tau(\ket{\psi_0(\phi,\varphi_\textrm{s})})=\tau_{\min}(e^{-iH t/\hbar}\ket{\psi_0(\phi,\varphi_\textrm{s})})\equiv \bar\tau.
\eeq
In this way, when $U_\textrm{s}(\varphi_\textrm{s}-\varphi_\textrm{opt})$ is applied at a time $\Delta \tilde t$ after the maximal 3-tangle has been reached, the entanglement exhibits the oscillatory behaviour depicted in Fig. \ref{fig:evol-libre}.  
If $\Delta \tilde t$ is small enough such that $\bar\tau\geq\tau^\star$, this protocol maintains the 3-tangle above the threshold value $\tau^{\star}$, as shown in all curves of Figure \ref{fig:evol-libre}.

Figure \ref{fig:optima-magica-libre} compares the evolution of $\tau$ when different protocols are implemented on the same initial state $\ket{\psi_0}$. 
Firstly, a change to the optimal relative phase is made at $t=0$. The resulting state $\ket{\psi^{\textrm{op}}_0}$ evolves under the action of the Hamiltonian until $\tilde t=\tilde t_{1}+\delta \tilde t$, when three different operations are implemented: 
the state's relative phase is adjusted to its optimal value at time intervals $\delta \tilde t=\tilde t_2-\tilde t_1$, following the protocol (\ref{prot_gelatina})  (solid green curve); 
a rotation $R_{\textrm{s}}$ is successively applied at time intervals $\delta\tilde t$, and the state rapidly approximates to $\ket{\psi_{\textrm{s}}}$ (red-dashed curve); 
a single rotation $R_{\textrm{s}}$ is performed and afterwards the system is let to evolve solely under the action of the Hamiltonian (blue-dotted curve).
Notice that whereas the blue dotted curve evolves freely (without the need of further additional operations besides the Hamiltonian evolution), the green solid line reaches higher values of $\tau$ more frequently, thus the system spends more time in highly-entangled states.    

\section{Summary and Final remarks}

We have considered  
the dynamical generation of three-way entanglement through the interaction among three qubits in a GHZ state, 
and proposed schemes to 
attain states with a high and sustained degree of entanglement in the shortest possible time.

States with maximal entanglement loss are those with maximal entanglement rate. 
A detailed examination of the latter revealed the optimal relative phase (between the states $\ket{000}$ and $\ket{111}$) that maximize the entanglement production at any stage of the evolution, for a fixed yet arbitrary amount of 3-tangle (fixed $\phi$) and a given Hamiltonian (fixed $r$).
This led to the optimization of $\Gamma$ via a one-qubit rotation $R_\textrm{op}$, an operation that locally transforms the actual state of the system into the (identically entangled) state that reaches the maximum amount of entanglement in the shortest possible time.

We further analyzed the effect of the flip operation $U_f$, and identified the conditions under which it improves the initial entanglement rate of a state, namely conditions   (\ref{cond_signo}) (provided $\Gamma_0<0$) and  (\ref{cond2}). 
In these cases, $U_f$ is useful in reducing the time needed to reach a highly entangled state by means of a local and universal operation. 
We also showed that the flip operation can be improved by performing the additional (state-dependent) rotation  
that shifts the actual relative phase to its optimal value. 
This two-steps procedure is equivalent to the single optimal rotation operation, thus giving two different paths to optimize $\Gamma$. 
The convenience of either one of these paths will depend on the specific setup and experimental capabilities.

Finally, we proposed different protocols aimed at maintaining the 3-tangle above some predetermined threshold value $\tau^*$ for all times. 
The first one is composed of a sequence of one-qubit operations: the rotation $R_\textrm{op}$ that shifts the relative phase to its optimal phase, followed by the successive application of the $\sigma_z$ operator (over either one of the qubits).
By adjusting the time $\delta \tilde t$ between consecutive applications of $\sigma_z$, it is possible to maintain the 3-tangle above the threshold value. 
If the sequence is applied with enough frequency, the tripartite entanglement  will remain sufficiently near to its maximum value $\tau=1$, a feature that  exhibits a resemblance to the quantum Zeno effect. 
Our second protocol avoids the need of repeated operations to prevent the 3-tangle from dropping below 
$\tau^*$, and rests on the identification of $\ket{\psi_s}$ as a steady state with maximal entanglement. 
The first step involves again the implementation of $R_\textrm{op}$, thus guaranteeing that the maximal entanglement is achieved in the fastest way. 
Once the state reaches the maximum entanglement, a second rotation $R_\textrm{s}$ is performed to set the relative phase to $\varphi_\textrm{s}$, and the subsequent evolving 3-tangle (result of the Hamiltonian evolution alone) oscillates above a certain value that can be adjusted by changing the time at which $R_\textrm{s}$ was implemented.

Our analysis contributes to the design of protocols aimed at speeding-up the production of three-way entanglement by means of a non-local Hamiltonian assisted by local ---in most cases one-qubit--- operations, and paves the way for future analysis regarding the optimization of \emph{multipartite} entanglement rate in more general composite systems.  


\begin{acknowledgments}
N.G. and A.P.M. acknowledge funding from Grants
No. PICT 2020-SERIEA-00959 from ANPCyT (Argentina) and No. PIP 11220210100963CO from CONICET (Argentina). A.P.M. acknowledges partial support from SeCyT, Universidad Nacional de Córdoba (UNC), Argentina.
A.V.H acknowledges financial support from DGAPA, UNAM through project PAPIIT IN112723. L.H.M acknowledges financial support of CONAHCyT through fellowship 
812988. 
\end{acknowledgments}

\bibliographystyle{iopart-num}
\bibliography{Entanglement-rate_GHZ}{}

\end{document}